\documentclass[twocolumn,floatfix,amsmath,amssymb,prl, superscriptaddress,10pt]{revtex4}

\usepackage{epsfig}
\usepackage{graphicx}
\usepackage{dcolumn}
\usepackage{bm}
\usepackage{natbib}

\begin{document}

\title{Nanometer-scale spatial modulation of an inter-atomic interaction in a Bose-Einstein condensate}
\author{Rekishu Yamazaki}
\affiliation{Graduate School of Science, Kyoto University, Kitashirakawa Oiwake-cho, Sakyo-ku, Kyoto, 606-8502, Japan}
\affiliation{JST-CREST, 4-1-8 Honmachi, Kawaguchi, Saitama 331-0012, Japan}
\author{Shintaro Taie}
\author{Seiji Sugawa}
\affiliation{Graduate School of Science, Kyoto University, Kitashirakawa Oiwake-cho, Sakyo-ku, Kyoto, 606-8502, Japan}
\author{Yoshiro Takahashi}
\affiliation{Graduate School of Science, Kyoto University, Kitashirakawa Oiwake-cho, Sakyo-ku, Kyoto, 606-8502, Japan}
\affiliation{JST-CREST, 4-1-8 Honmachi, Kawaguchi, Saitama 331-0012, Japan}
\date{\today}

\begin{abstract}
We demonstrate nanometer-scale spatial control of inter-atomic interactions in a Bose-Einstein condensate of ytterbium(Yb).  A pulsed optical standing wave, tuned near an optical Feshbach resonance varies the $s$-wave scattering length continuously across the standing wave pattern. The modulated mean-field energy with a spatial period of every 278 nm is monitored by a diffraction pattern in a time-of-flight image.  We observe a wide scattering length control of up to 160 nm. The demonstrated spatial modulation of the scattering length proves that the high resolution control of atomic interactions is possible.
\end{abstract}

\maketitle

Tunability of interatomic interactions using Feshbach resonance methods \cite{Mfeshbach} has opened up a variety of possibilities in the study of ultracold quantum gases such as formation of ultracold molecules \cite{coldmolecule}, a Bose-Einstein condensate (BEC) to Bardeen-Cooper-Schrieffer crossover with fermionic gases \cite{BECBCS}, simulation of super-nova (Bose-nova) \cite{bosenova}, and Efimov trimer states \cite{Efimov}.
 In these studies, the Feshbach resonances are induced by magnetically tuning the hyperfine energy level of a bound state to the vicinity of the incoming scattering state.  The length scale for application of the Feshbach field has, so far, been much larger than the size of the atomic sample.

\begin{figure}[b]
  \includegraphics[width=8cm]{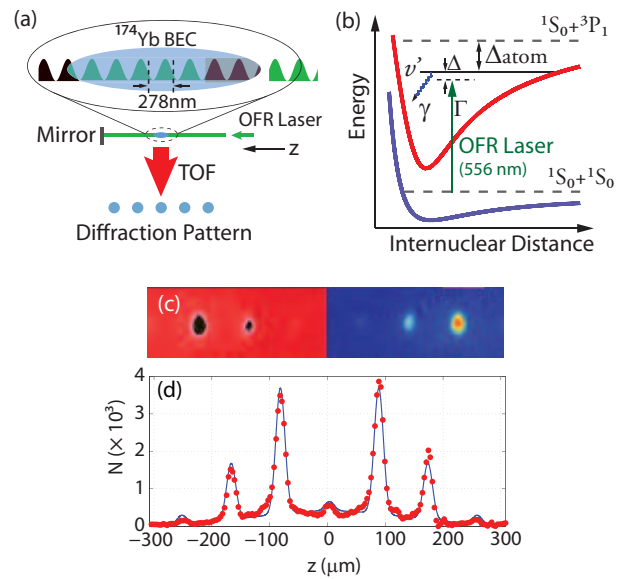}\\
  \caption{(color online).  (a) Schematic of the experimental setup.  A $^{174}$Yb condensate is irradiated with the standing wave formed by an OFR laser.  The diffraction pattern in the time-of-flight (TOF) image is observed.  (b) Energy diagram of the relevant states for the experiment.  The $^1$S$_0$-$^3$P$_1$ photoassociation transitions to the vibrational states $v'$ = 11, 12, and 13 are used for the OFR.  (c) Typical diffraction pattern obtained in the experiment.  Each peak in the image represents different momentum components imparted by the pulsed lattice.  (d) An integrated column density of the image where the red dots and a blue line corresponds to the data and the fitted line for the determination of $\beta$, respectively.  }\label{fig1}
\end{figure}

Many interesting experiments such as a sonic-analogue of black holes \cite{blackhole1}, quantum simulation of novel Hubbard model \cite{hubbard2, hubbard1}, and unusual Bose-nova phenomena \cite{bosenova2} could be explored with spatial modulation of the inter-atomic interaction on short length scales.
An alternative approach, an \emph{optical Feshbach resonance} (OFR), utilizing optical coupling between bound and scattering states \cite{OFRTheory}, is a promising technique for providing this possibility.  With OFR methods, the intensity and the detuning of the coupling laser are used to control the scattering length \cite{OFRExp}.  Fine modulation of laser intensity using optical standing wave and holographic technique is routinely shown in other applications \cite{1DLattice, hologram}, and these techniques combined with OFR should realize the fine spatial modulation of the scattering length.  Inherent nature of the optical control also enables fast manipulation and, moreover, optical transitions can be found in almost any states and the control is possible for numerous situations and atomic species.

Important progress along this direction has been reported in Ref.\,\cite{MOFR1} where successful control of a magnetic Feshbach resonance of alkali-metal atoms with a laser light was demonstrated.  The fine spatial control of the scattering length has, however, not yet been demonstrated, so far.  In this Letter, we report our successful demonstration of OFR modulation of the \emph{s}-wave scattering lengths on the scale of hundreds of nm.

For the demonstration of submicron control of the scattering length, we apply a pulsed optical standing wave to a BEC of ytterbium ($^{174}$Yb) atoms.  The pulsed light for OFR is tuned to the vicinity of the weak intercombination $^1$S$_0$-$^3$P$_1$ photoassociation (PA) resonances with the resonant wavelength of $556$ nm, which results in a modulation of the mean-field energy of the condensate with a spatial period of $278$ nm. The resulting phase modulation of the condensate is observed in the diffraction pattern in a time-of-flight (TOF) images.  The observed behavior is well explained by the semi-classical theory of Bohn and Julienne \cite{PLOFRAnalysis}. Our analysis shows scattering length variation of up to $160$ nm over a comparable distance of only 280nm.

To determine the expected change of scattering length $a$ and PA rate $K$ due to the applied laser field, the formalism of Bohn and Julienne is valid under our conditions.  In their formalism, both parameters can be calculated in terms of the radiative decay rate of molecule $\gamma$, the light-induced width $\Gamma$ and the detuning of the OFR laser from a  photoassociation resonance $\Delta$.  The energy level diagrams relevant to this work is shown in Fig.\,$1$(b).  The light-induced width $\Gamma$ is given by $\Gamma(I)=3I\gamma\lambda^3f_{rot}f_{FC}/8\pi c$, where $f_{rot}$ = 1/3 is the rotational factor for the relevant transition, $f_{FC}$ is the Frank-Condon factor, $c$ is the speed of light in a vacuum, and $\lambda=555.8$ nm and $I$ are the wavelength and the intensity of the OFR laser, respectively.  For the weak excitation regime, where $\Gamma\ll\gamma$, the scattering parameters are given by
\begin{eqnarray}
a&=&a_{bg}+\delta a\nonumber\\
 &=&a_{bg}+\frac{1}{2k}\frac{\Gamma\Delta}{\Delta^2+\gamma^2/4} \label{eqa}\\
K&=&\frac{\pi\hbar}{\mu k} \frac{\Gamma\gamma}{\Delta^2+\gamma^2/4 \label{eqK}},
\end{eqnarray}
where $a_{bg}=5.55$ nm is the background $s$-wave scattering length of $^{174}$Yb, $\mu=m/2$ is the reduced mass of two scattering atoms with mass $m$, and $k=\sqrt{\frac{21}{8}}\frac{1}{2R_{TF}}$ is the wave number calculated from the momentum of the condensate with Thomas-Fermi radius $R_{TF}$ \cite{Bragg}.  $\delta a$ is the variation of the scattering length due to the OFR and is proportional to $I$.

Despite its potentially wide applicability, the use of OFR has been scant \cite{OFRExp}. A reason for this is that usually the optical coupling also induces inelastic scattering, leading to rapid atom depletion.  A use of narrow transitions in alkaline-earth-metal-like atoms \cite{OFRYbTh} to avoid inelastic scattering loss was successfully demonstrated in a recent experiment in our group using the $^1$S$_0$-$^3$P$_1$ intercombination transition in thermal gases of $^{172}$Yb and $^{176}$Yb \cite{OFRYbExp}.

In that earlier work a relatively low inelastic scattering rate was observed, an order of magnitude less than the case of alkaline atoms indicating the possibility of observing large tunings of scattering length over short distances, as we report here.  In this work we extend this technique to a condensate of $^{174}$Yb.

To demonstrate the capability of OFR to modulate the $s$-wave scattering length on short length scale and also on short time scale, we use a pulsed optical lattice beam as the OFR light, as shown in Fig.\,$1$(a).  The application of a pulsed optical lattice beam generally results in a diffraction of a released BEC, which is previously studied both theoretically and experimentally \cite{1DLattice}.  In a phase modulation regime, where the exposure time of the optical lattice $\tau$ is much smaller than the minimum classical oscillation period of the formed lattice, the effect of the lattice is treated as a thin grating.  The phase of the condensate modified by the lattice has a form
\begin{equation}
    \phi(z)=\frac{U(z)\tau}{\hbar}
\end{equation}
with $U(z)=U_0\cos^2(qz)$, where $q$ is the wavenumber of the lattice laser and $z$ is the direction along the lattice laser propagation.
The potential  $U_0$ experienced by the atoms in the ground state at an anti-node of the optical lattice is given by

\begin{equation}\label{eqUOFR}
    U_0=\frac{\hbar\Omega^2}{\Delta_{atom}} + \frac{4\pi\hbar^2n_a}{m} \delta a
\end{equation}

The first term describes the atomic light shift, where $\Omega$ and $\Delta_{atom}$ corresponds to the Rabi frequency and the detuning of the OFR laser to the atomic excited state, respectively.  The second term represents the OFR-induced shift of the mean-field energy, and thus is proportional to the scattering length variation $\delta a$ and the atom density $n_a$.  It is noted that there is a mean-field energy $U_{MF}=\frac{4\pi\hbar^2n_a}{m} a_{bg}$ across the condensate.  However, it is not susceptible to the OFR laser and does not contribute to the diffraction pattern.  It is clear from Eq.(\ref{eqa}) and (\ref{eqUOFR}), the OFR dispersively varies the scattering length, and therefore the mean-field energy, across a photoassociation resonance.  From the diffraction pattern of the condensate generated from the phase modulation $\phi(z)$ imparted by the OFR, one should be able to extract the variation of the scattering length.

The method for the all-optical formation of $^{174}$Yb condensate is described in Ref.\,\cite{BECYb}.
After the evaporation in a crossed far-off resonant trap (FORT), an almost pure $^{174}$Yb condensate is prepared with an atom number of up to 1.5$\times$10$^5$ in a typical harmonic trap potential $\bar{\omega}=(\omega_x \omega_y \omega_z)^{1/3}=2\pi\times 130$ Hz.  A simple schematic of the experimental process after the preparation of condensate is shown in Fig.\,\ref{fig1}(a).  Following the condensate formation, we release the condensate from the trap by turning off the FORT lasers.
At the release, the OFR laser pulse of a 1D optical lattice is turned on for a several microseconds with a typical power of 1-100 $\mu$W.  The OFR laser is tuned near the $^1$S$_0$-$^3$P$_1$ photoassociation resonances with the vibrational quantum number $v'$ = 11, 12 and 13, which correspond to the detuning $\Delta_{atom}$ of  -69, -117, -192 MHz, respectively \cite{resposition}.  After a TOF-time of typically 10 ms, the absorption image is taken for the diffraction pattern analysis.  For every OFR pulse, the power of the pulse is monitored by a fast photodiode and recorded with an oscilloscope to compensate for the power instability.

A typical image and a column density of the obtained diffraction pattern are shown in Fig.\,\ref{fig1}(c) and \ref{fig1}(d), respectively.  Each peak in the image represents the momentum component $p_n=n2\hbar q$, ($n=0,\pm1,\pm2,\ldots$) imparted by the pulsed lattice.  Population distribution for each momentum state is described as $P_n=J^2_n(\beta)$, where $\beta=U_0\tau/2\hbar$ and $J_n(z)$ are Bessel functions of the first kind \cite{1DLattice}.  For the image analysis, we first sum the image along the vertical axis to obtain an integrated column density profile in the horizontal axis, and fit the density profile by the Bessel functions to obtain the modulation index $\beta$.
Each peak is convoluted with a Gaussian function to account for the momentum spread and the finite image resolution.
It is also noted that we subtract the background distribution in the images.  There are two origins of the background distribution; the residual thermal component and the elastic scattering of wave packets of condensate with different momentum during the TOF \cite{InElastic1,InElastic2}.  Strictly speaking, the atom distribution from the later phenomena are non-Gaussian. However, due to low atom counts from the elastic scattering we do not observe clear deviation from the Gaussian fit for the background distribution.

\begin{figure}
  \includegraphics[width=8.5cm]{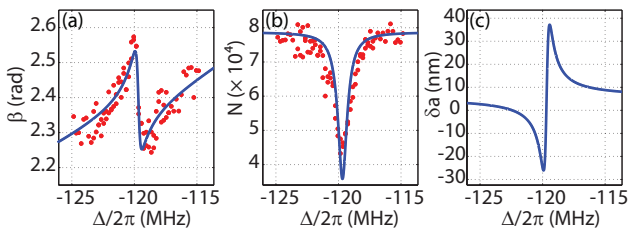}\\
  \caption{(color online).  (a) Observed modulation index $\beta$, (b) number of atoms remained after the OFR pulse, and (c) calculated scattering length for the $v'=12$ transition.     We first fit the variation of $\beta$ using the formalism by Bohn and Julienne to determine light-induced width $\Gamma$.  Blue lines in (b) and (c) are calculated atom loss and scattering length variation using the $\Gamma$ obtained from the fit in (a), respectively. }\label{fig2}
\end{figure}

The images of the diffraction pattern  across the photoassociation resonances are collected by scanning the OFR laser frequency. We keep the OFR pulse length to be short and all the diffraction pattern images collected are in the phase modulation regime. Shown in Fig.\,\ref{fig2} are typical $\beta$ and atom number after the OFR pulse as obtained from the images.  The observed $\beta$ has a dispersive shape centered at the photoassociation resonance, where atom loss is maximum.  The slight increase in $\beta$ towards the atomic resonance is due to the variation of the atomic light shift.  To extract the scattering length from the data,  we first fit the observed $\beta$ to determine $\Gamma$ and $\Omega$.  We then calculate the atom loss and the variation of the scattering length $\delta a$ from these parameters.  Atom loss obtained from the data agrees well with that obtained from the fit to the $\beta$ variation.

\begin{figure}
  \includegraphics[width=8.5cm]{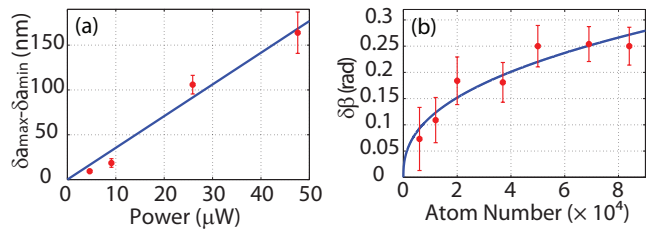}\\
  \caption{(color online).  (a) Power dependence of the scattering length variation along with a linear fit line.  The linear dependence is consistent with the weak excitation expected from the $\Gamma$ value obtained from the data.  (b) Atom number dependence of $\delta\beta$ along with the best fitted curve in blue.  The OFR induced mean-field variation is expected to depends on the atom density which is proportional to $N^{2/5}$ in a harmonic potential.  From the fit, we obtained $p=0.41$ for $N^{p}$ dependence fitting, which agrees well with the predicted density dependence.  The error bars in both figures show the root mean squared error of the data obtained from the fitting procedure of the modulation index $\beta$. }\label{fig3}
\end{figure}

The clear dispersive variation in $\beta$ is direct evidence of the OFR effect.  We further investigate the behavior of the signals to confirm the OFR effect.  First, the variation of $\delta a$ with respect to the OFR laser power is measured.
The fact that $\Gamma/2\pi<20$ kHz  for all the data obtained means that all the measurements are performed in the weak excitation condition $\Gamma\ll\gamma$ (=2$\pi\times$364 kHz) where $\delta a$ should be proportional to the laser intensity as shown in Eq.\,(1).  The results shown in Fig.\,\ref{fig3}(a) clearly confirm the linear dependence, which is consistent with the weak excitation regime.  The observed variation of the scattering length up to 160 nm is quite large.
It is important to note that $\delta a$ varies continuously from the anti-node, where the effect of OFR is maximum, to the node where the effect of OFR is absent.
The OFR laser wavelength of $\lambda=555.8$ nm results in a periodic modulation of $\delta a$ in approximately 278 nm spacing.

Second, the variation of the modulation index across the photoassociation resonance $\delta\beta$ with respect to the atom number is measured.  The OFR effect is proportional to the density as is shown in Eq.\,(\ref{eqUOFR}).  For a given trap frequency, the density $n_a$ is proportional to $N^{2/5}$ \cite{pethicksmith}.  Figure\,\ref{fig3}(b) shows the variation of $\delta\beta$ as a function of the atom number with a best fit curve of $N^p$ shown with a blue line.  From the best fit, we obtain $p=0.41$ and the value agrees well with the $N^{2/5}$-dependence predicted.  The agreement between the data and the curve clearly shows that the observed $\delta\beta$ comes from the mean-field energy of the condensate.

\begin{figure}
  \includegraphics[width=5.5cm]{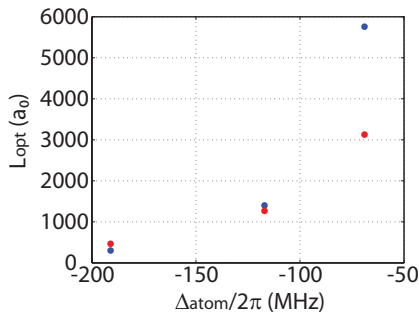}\\
  \caption{(color online).  Comparison between calculated (red dots) and measured (blue dots) $L_{opt}$.  The rapid increase of $L_{opt}$ for the shallower detuning are observed for both results.  The discrepancy between two results may be due to the uncertainty in the laser intensity estimation at the condensate. }\label{fig4}
\end{figure}

The effect of the OFR depends on the Frank-Condon factor of the photoassociation resonance chosen.  From the same measurement on different photoassociation resonances, we obtain the optical length defined as $L_{opt} = \Gamma/2\gamma$ which describes the strength of the OFR coupling for each resonance.  We plot the optical length obtained from the measurement (blue dots) and the numerical calculation (red dots) \cite{optL} in Fig.\,\ref{fig4}.  As the transition gets shallower toward the atomic resonance, the measured optical length shows faster increase in the optical length than the numerically calculated values. The tendency in the variation of $L_{opt}$, however, is consistent between two results.      Part of the discrepancy may be due to the uncertainty in the laser intensity estimation at the condensate.

In conclusion, we demonstrate rapid, nanometer-scale modulation of the scattering length using optical Feshbach resonances in a $^{174}$Yb condensate.  A large scattering length variation of up to 160 nm is obtained with a spatial modulation of 278 nm by observing a diffraction pattern in a time-of-flight image.  The demonstrated fine spatial modulation of the scattering length opens up a wide variety of applications.  New applications may benefit not only from the fine spatial resolution offered by a OFR, but also the inherently available fast manipulation \cite{fastManip1}.  The time constant for the manipulation of the OFR can be quite fast compared to the magnetic Feshbach resonances and could be used to control and study the non-equilibrium dynamics of the condensate.  In addition to these possibilities, creation of multiple frequency super-lattices with frequency components tuned to different OFR resonances may create intricate patterns of large variation in the interaction strength. Independent control of different scattering lengths \cite{multiMunip}, and local manipulation of the condensate or the condensate in the lattices may also be possible.

We acknowledge S. Uetake for the experimental help. This work is supported by the Grant-in-Aid for Scientific Research of JSPS (No.18204035, 21102005C01, 21104513A03 Quantum Cybernetics), GCOE Program ``The Next Generation of Physics, Spun from Universality and Emergence" from MEXT of Japan, and World-Leading Innovative R\&D on Science and Technology (FIRST). ST and SS acknowledge supports from JSPS.

\end{document}